%% file: gyre-tides.tex
\documentclass[twocolumn]{aastex631}

\usepackage{amsmath}
\usepackage{amssymb}
\usepackage{natbib}
\usepackage{CJKutf8}
\usepackage{enumitem}

\usepackage{savesym}
\savesymbol{tablenum}
\usepackage{siunitx}
\input{symbols}

\graphicspath{{./}{figures/}}

\begin{document}

\title{\xgyretides: Modeling binary tides within the \gyre\ stellar oscillation code}

\begin{CJK*}{UTF8}{gbsn}
\author[0000-0001-9037-6180]{Meng Sun (孙萌)}
\affiliation{Department of Astronomy, University of Wisconsin-Madison, 475 N Charter St, Madison, WI 53706, USA}
\affiliation{Center for Interdisciplinary Exploration and Research in Astrophysics (CIERA), Northwestern University, 1800 Sherman Ave, Evanston, IL 60201, USA}
\author[0000-0002-2522-8605]{R. H. D. Townsend}
\affiliation{Department of Astronomy, University of Wisconsin-Madison, 475 N Charter St, Madison, WI 53706, USA}
\author[0000-0002-0951-2171]{Zhao Guo}
\affiliation{Department of Applied Mathematics and Theoretical Physics (DAMTP), University of Cambridge, Cambridge CB3 0WA, UK}
\affiliation{Center for High Angular Resolution Astronomy and Department of Physics and Astronomy, Georgia State University, Atlanta, GA, USA}
\email{meng.sun@northwestern.edu}
\email{townsend@astro.wisc.edu}
\email{zg281@cam.ac.uk}

\begin{abstract}
We describe new functionality in the \gyre\ stellar oscillation code
for modeling tides in binary systems. Using a multipolar expansion in
space and a Fourier-series expansion in time, we decompose the tidal
potential into a superposition of partial tidal potentials. The
equations governing the small-amplitude response of a
spherical star to an individual partial potential are the
linear, non-radial, non-adiabatic oscillation equations with an extra
inhomogeneous forcing term. We introduce a new executable,
\xgyretides, that directly solves these equations within the
\gyre\ numerical framework. Applying this to selected problems, we
find general agreement with results in the published literature but
also uncover some differences between our direct solution methodology
and the modal decomposition approach adopted by many authors.

In its present form \xgyretides\ can model equilibrium and dynamical
tides of aligned binaries in which radiative diffusion dominates the
tidal dissipation (typically, intermediate and high-mass stars on the
main sequence). Milestones for future development include
incorporation of other dissipation processes, spin-orbit misalignment,
and the Coriolis force arising from rotation.
\end{abstract}


\section{Introduction} \label{s:intro}

The \gyre\ stellar oscillation code
\citep{Townsend:2013,Townsend:2018,Goldstein:2020} is a open-source
software instrument that solves the linear, non-radial, non-adiabatic
oscillation equations for an input stellar model. Released in 2013, it
has been productively used to study of heat-driven oscillations in
$\gamma$ Doradus and $\delta$ Scuti pulsators
\citep[e.g.,][]{Van-Reeth:2022,Murphy:2022}, slowly-pulsating B stars
\citep[e.g.,][]{Michielsen:2021}, variable sub-dwarf B stars
\citep[e.g.][]{Silvotti:2022}, pulsating pre-main sequence stars
\citep[e.g.,][]{Steindl:2021}, DBV white dwarfs
\citep[e.g.,][]{Chidester:2021}, and hypothetical `dark' stars
\citep{Rindler-Daller:2021}; to explore stochastically excited
oscillations in solar-like, subgiant and red-giant stars
\citep[e.g.,][]{Nsamba:2021,LiTanda:2020,LiTanda:2022}; and to model
oscillations of uncertain origin in red supergiant stars
\citep{Goldberg:2020}, post-outburst recurrent novae \citep{Wolf:2018}
and even gas-giant planets \citep{Mankovich:2019}.

This paper describes new functionality in \gyre\ for modeling static
and dynamic tides in binary\footnote{While our narrative focuses on
binary \emph{star} systems, it remains equally applicable to
star-planet systems.} systems. The equations governing small tidal
perturbations to one component of a binary are the linear
oscillation equations with extra terms representing the gravitational
forcing by the companion. Release 7.0 of \gyre\ implements these terms
and the supporting infrastructure necessary to solve the tidal
equations.

The view of astrophysical tides through the lens of forced
oscillations was pioneered in a pair of seminal papers by
\citet{Zahn:1970,Zahn:1975}. These papers also introduce complementary
approaches to solving the tidal equations, that we dub `modal
decomposition' (MD) and `direct solution' (DS). In MD the tidal
perturbations are decomposed as a superposition of the star's
free-oscillation modes, with weights determined from overlap integrals
between the mode eigenfunctions and the tidal force field. In DS the
two-point boundary value problem (BVP) posed by the tidal equations is
solved directly using a standard approach such as shooting or
relaxation. Examples of MD are given by \cite{Kumar:1995},
\citet{Lai:1997}, \citet{Schenk:2001}, \citet{Arras:2003} and \citet{Fuller:2012}; and of DS by
\citet{Savonije:1983,Savonije:1984}, \citet{Pfahl:2008}, and
\citet{Valsecchi:2013}. The study by \citet[][hereafter
  B12]{Burkart:2012} is noteworthy in that it adopts \emph{both}
approaches, although no direct comparison is made between them (a
lacuna that appears to extend into the wider literature).

The following section lays out the theoretical foundations for our
treatment of tides. Section~\ref{s:implement} describes the
modifications to \gyre\ to implement this formalism via a DS
methodology, and Section~\ref{s:calcs} presents illustrative calculations
focused on selected problems in the published
literature. Section~\ref{s:discuss} summarizes the paper, discusses
potential applications of the new \gyre\ functionality, and outlines
future improvements.


\section{Theoretical Formalism} \label{s:formalism}

Rather an exhaustive derivation of all equations, we opt to focus on
the key expressions that support our narrative and define the choices
(e.g., normalizations, sign conventions) dictated by the existing
numerical framework of \gyre. For more-detailed exposition,
we refer the reader to the papers by \citet{Polfliet:1990},
\citet{Smeyers:1991}, \citet{Smeyers:1998}, and
\citet{Willems:2003,Willems:2010}.

\subsection{Binary Configuration} \label{s:formalism-config}

Consider a binary system comprising a primary star of mass $\Mpri$
and photospheric radius $\Rpri$, together with a secondary star of
mass $\mratio \Mpri$. To model the tides raised on the primary by the
secondary, we adopt a non-rotating reference frame with the primary's
center-of-mass fixed at the origin, the orbit of the secondary lying
in the Cartesian $xy$-plane, and the line of apsides coinciding with
the $x$-axis. The position vector of the secondary at time $t$ is then
given by
\begin{equation} \label{e:vsec}
  \vsec(t) = \frac{a(1 - e^{2})}{1 + e \cos\tanom} \left( \vex \cos
  \tanom + \vey \sin \tanom \right),
\end{equation}
where $\vex$ and $\vey$ are the unit vectors along the $x$ and $y$
axes, respectively, $a$ is the orbital semi-major axis, and $e$ the
eccentricity. The true anomaly $\tanom$ is linked to $t$ via Kepler's
equation
\begin{equation} \label{e:kepler}
  \eanom - e \sin\eanom = \manom
\end{equation}
and the auxiliary relations
\begin{gather}
  \manom = \Oorb (t - \tperi), \\
  \tan \frac{\tanom}{2} = \sqrt{\frac{1-e}{1+e}} \tan \frac{\eanom}{2},
\end{gather}
where $\tperi$ is a time of periastron passage, and $\manom$ and
$\eanom$ are the mean and eccentric anomalies, respectively.  The
orbital angular frequency $\Oorb$ is given by Kepler's third law,
\begin{equation}
  G \Mpri (1 + \mratio) = a^{3} \Oorb^{2},
\end{equation}
with $G$ the gravitational constant.

\subsection{Tidal Potential} \label{s:formalism-pot}

Tides are raised on the primary star by the forces arising from the
gravitational potential $\potsec$ of the secondary, which at position
vector $\vpos$ and time $t$ is
\begin{equation}
  \potsec(\vpos;t) = - \frac{\mratio G \Mpri}{|\vsec - \vpos|}.
\end{equation}
Using a multipolar expansion in space and a Fourier-series expansion
in time, this expression can be recast as
\begin{multline} \label{e:pot-expand}
  \potsec(\vpos;t) =
  - \frac{\mratio G \Mpri}{\rsec}
  - \frac{\mratio G \Mpri}{\rsec^{2}} r \sin\vartheta \cos(\varphi - \tanom) + \mbox{} \\
  \pottide(\vpos;t)
  \end{multline}
where $(r,\vartheta,\varphi)$ are the spherical-polar radius,
colatitude and azimuth coordinates corresponding to $\vpos$. The first
term on the right-hand side is constant and therefore does not
generate a force. The second term produces a spatially uniform force
directed from the primary star toward the secondary, and precisely
cancels the inertial force arising from the orbital motion of the
primary center-of-mass about the system center-of-mass. The third term
represents the tidal part of the secondary potential, and is expressed
as a superposition
\begin{equation} \label{e:pottide}
  \pottide(\vpos;t) =
  \sum_{\ell=2}^{\infty} \sum_{m=-\ell}^{\ell} \sum_{k=-\infty}^{\infty}
  \pottide[\ell,m,k] (\vpos;t)
\end{equation}
of partial tidal potentials defined by
\begin{multline} \label{e:pottide-part}
  \pottide[\ell,m,k](\vpos;t) =
  - \epstide \,
  \frac{G \Mpri}{\Rpri} \,
  \cbar[\ell,m,k]
  \left( \frac{r}{\Rpri} \right)^{\ell} \times \mbox{} \\
  Y^{m}_{\ell}(\vartheta,\varphi) \,
  \ee{- \ii k \manom}.
\end{multline}
Here,
\begin{equation}
  \epstide \equiv
  \left( \frac{\Rpri}{a} \right)^{3} \, \mratio =
  \frac{\Oorb^{2} \Rpri^{3}}{G \Mpri} \, \frac{\mratio}{1 + \mratio}
\end{equation}
is a dimensionless parameter that quantifies the overall strength of
the tidal forcing, $\cbar[\ell,m,k]$ is an tidal expansion coefficient
(Appendix~\ref{a:exp-coeff}), and $Y^{m}_{\ell}$ is a spherical
harmonic (Appendix~\ref{a:sph-harm}).

\subsection{Tidal Response} \label{s:formalism-resp}

Appendix~\ref{a:lin-eqns} summarizes the set of linearized equations
governing the response of the primary star to the tidal potential
$\pottide$. Based on the form~(\ref{e:pottide-part}) of the partial
potentials, solutions to these equations take the form
\begin{multline} \label{e:sol-xi}
  \vxi(\vpos;t) =
  \sum_{\ell,m,k}
  \left[
    \txir[\ell,m,k](r) \, \ver + \phantom{\left( \vet \, \pderiv{}{\vartheta} \right)}
  \right. \\
  \left.
    \txih[\ell,m,k](r) \left( \vet \, \pderiv{}{\vartheta} + \frac{\vep}{\sin\vartheta} \, \pderiv{}{\varphi}\right)
  \right] \,
  \angtime[\ell,m,k](\vartheta,\varphi;t)
\end{multline}
for the displacement perturbation vector $\vxi$, and
\begin{equation} \label{e:sol-f}
  f'(\vpos;t) =
  \sum_{\ell,m,k}
  \tf[\ell,m,k]'(r) \, \angtime(\vartheta,\varphi;t)
\end{equation}
for the Eulerian ($f'$) perturbation to a scalar variable $f$ (the
corresponding Lagrangian perturbation $\delta f$ follows from
equation~\ref{e:lag-eul}). In these expressions, the notation
$\sum_{\ell,m,k}$ abbreviates the triple sum appearing in
equation~(\ref{e:pottide}), while $\ver$, $\vet$ and $\vep$ are the
unit basis vectors in the radial, polar, and azimuthal directions,
respectively. The functions
\begin{equation} \label{e:angtime}
  \angtime[\ell,m,k](\vartheta,\varphi;t) \equiv Y^{m}_{\ell}(\vartheta,\varphi) \ee{-\ii k \manom}
\end{equation}
describe the angular and time dependence of the response, while the
functions with tilde accents ($\txir,\txih,\tf'$) encapsulate the
radial dependence. The latter are found as solutions to a system of
tidal equations summarized in
Appendix~\ref{a:tidal-eqns}. Importantly, the set of radial functions
for a given combination of indices $\{\ell,m,k\}$ can be determined
independently of any other combination.

\section{Implementation in \gyre} \label{s:implement}

To implement the tidal equations~(\ref{e:tidal-mass}--\ref{e:tidal-diff}) in \gyre, which follows a DS methodology, we first
transform to a dimensionless independent variable $x \equiv r/\Rpri$
and a set of dimensionless dependent variables
\begin{equation}
  \begin{aligned}
    y_{1} &\equiv x^{2 - \ell} \frac{\txir}{r}, \\
    y_{2} &\equiv x^{2 - \ell} \frac{\tP'}{\rho g r}, \\
    y_{3} &\equiv x^{2 - \ell} \frac{\tpottot'}{gr}, \\
    y_{4} &\equiv x^{2 - \ell} \frac{1}{g} \deriv{\tpottot'}{r}, \\
    y_{5} &\equiv x^{2 - \ell} \frac{\delta \tS}{\cP}, \\
    y_{6} &\equiv x^{-1 - \ell} \frac{\delta \tLrad}{\Lpri}.
  \end{aligned}
\end{equation}
Here $g \equiv \sderiv{\potpri}{r}$ is the scalar gravity, and the other symbols are defined in Appendix~\ref{a:lin-eqns} (for notational simplicity we neglect the $\ell,m,k$ subscripts from perturbed quantities). With these transformations, the differential equations and boundary conditions
governing $y_{1},\ldots,y_{6}$ can be written in a form almost
identical to the linear, non-radial, non-adiabatic free oscillation equations detailed in
Appendix B2 of \citet{Townsend:2018}. The only differences are that
the interpretation of the $y_{3}$ and $y_{4}$ variables is altered;
the outer boundary condition governing the gravitational potential
acquires an inhomogeneous term on the right-hand side
\begin{equation} \label{e:inhom-bc}
 U y_{1} +  (\ell + 1) y_{3} + y_{4} = (2 \ell + 1) y_{\rm T},
\end{equation}
where $U$ is the usual homology invariant \citep[e.g.,][]{Kippenhahn:2013} and
\begin{equation} \label{e:y-T}
  y_{\rm T} \equiv x^{2 - \ell} \frac{\tpottide[\ell,m,k]}{gr};
\end{equation}
and the dimensionless oscillation frequency is defined by
\begin{equation} \label{e:omega}
  \omega \equiv \sqrt{\frac{\Rpri^{3}}{G \Mpri}} \sigmk,
\end{equation}
where $\sigmk$ is the rotating-frame frequency defined in
equation~(\ref{e:sigmk}).

\gyre\ uses a multiple shooting algorithm to discretize BVPs on a
radial grid $x = x_{1}, \ldots, x_{N}$. For a system of of $n$
differential equations and $n$ boundary conditions (in the present
case, $n=6$) this leads to a corresponding system of linear algebraic
equations with the form
\begin{equation} \label{e:lin-sys}
  \mS \vu = \vb,
\end{equation}
where $\mS \in \mathbb{C}^{N n \times N n}$ is a block staircase
matrix and the solution vector $\vu \in \mathbb{C}^{N n}$ contains the
dependent variables $y_{1},\ldots,y_{n}$ evaluated at successive grid
points. For free-oscillation problems the right-hand side vector $\vb
\in \mathbb{C}^{N n}$ is identically zero, and so the linear
system~(\ref{e:lin-sys}) is homogeneous; non-trivial solutions exist
only when the determinant of $\mS$ vanishes, a condition that
ultimately determines the dimensionless eigenfrequencies $\omega$ of
the star. However, for the tidal problem considered here, $\vb$ is
non-zero due to the inhomogeneous term appearing on the right-hand
side of the boundary condition~(\ref{e:inhom-bc}). The linear system
can then be solved for any choice of $\omega$ --- that is, the
dimensionless frequency is an \emph{input} rather than an output.

The reordered flow of execution ($\omega$ as input rather than output)
motivates the decision to provide two separate executables in release
7.0 of \gyre: \xgyre\ for modeling free stellar oscillations (the same
as in previous releases), and \xgyretides\ for modeling stellar tides
(new to this release). A full description of these programs, including
their input parameters and output data, is provided on the
\gyre\ documentation
site\footnote{\url{https://gyre.readthedocs.io/en/stable/}}


\section{Example Calculations} \label{s:calcs}

\subsection{Surface Perturbations in a KOI-54 Model} \label{s:calcs-surf}

\begin{table}
  \begin{centering}
  \begin{tabular}{cccccc}
    Model  & $M$ (\si{\Msun}) & $R$ (\si{\Rsun}) & $\Teff$ (\si{\kelvin}) & $Z_{\rm i}$ & $X_{\rm c}$ \\ \hline
    KOI-54 & 2.32 & 2.19 & \num{9400}  & 0.0328 & 0.487 \\
    B-star & 5.00 & 2.80 & \num{16700} & 0.0200 & 0.658 \\
  \end{tabular}
  \caption{Fundamental parameters for the two stellar models
    discussed in the text: mass $M$, radius $R$, effective
    temperature $\Teff$, initial metal mass fraction $Z_{\rm i}$ and center hydrogen mass fraction $X_{\rm c}$.} \label{t:models}
  \end{centering}
\end{table}

\begin{figure}
  \includegraphics{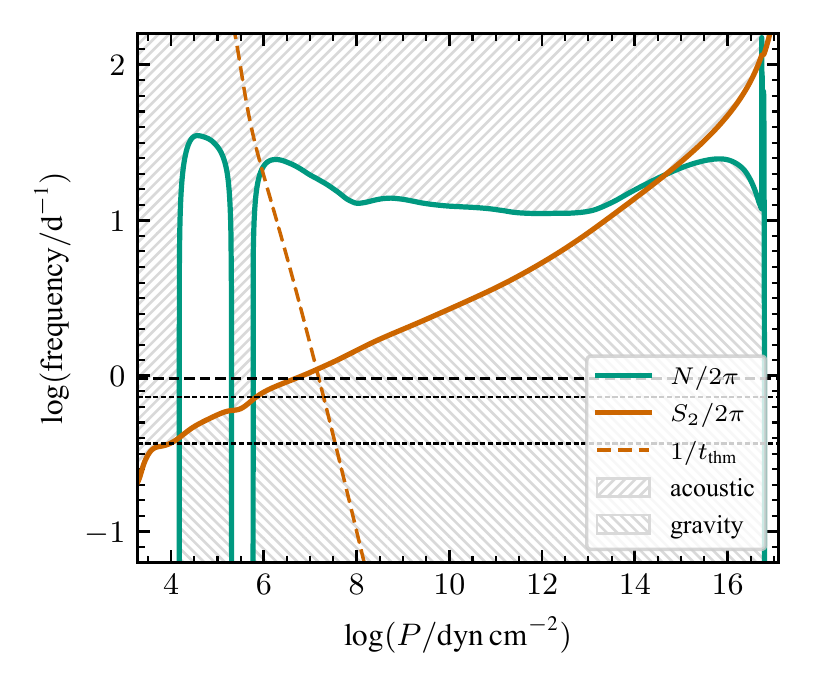}
  \caption{Propagation diagram for the KOI-54 model discussed in the
    text, plotting the \BV\ ($N$) and Lamb ($\Slamb$) frequencies as a
    function of pressure coordinate for $\ell=2$ modes. The hatched
    regions indicate the acoustic wave ($\sigma > N,\Slamb$) and
    internal gravity wave q($\sigma < N,\Slamb$) propagation
    regions. The sloped dashed line shows the frequency corresponding
to the local thermal timescale $\tthm$ (equation~\ref{e:t-thm}), while the upper (lower) horizontal dashed line corresponds to a frequency $\sigma/\Oorb = 30$ (15).}
  \label{f:prop-diag-koi54}
\end{figure}

KOI-54 is a highly-eccentric ($e \approx 0.8$), near face-on binary
system comprising a pair of A-type stars. Discovered by
\citet{Welsh:2011} in Kepler observations, it serves as the archetype
of the heartbeat class of periodic variables
\citep{Thompson:2012}. B12 and \citet{Fuller:2012} each present
initial attempts to model the light curve of KOI-54 in terms of
contributions from the equilibrium tide, dynamical tides and stellar
irradiation. Here we undertake a calculation to reproduce Fig.~6 of
B12, which illustrates how a model for the KOI-54 primary responds to
forcing by a single partial tidal potential.

Guided by the parameters given in Table~1 of B12, we use release
r22.11.1 of the \mesa\ software instrument
\citep{Paxton:2011,Paxton:2013,Paxton:2015,Paxton:2018,Paxton:2019,Jermyn:2022}
to evolve a \SI{2.32}{\Msun} model from the zero-age main sequence
(ZAMS) until its photospheric radius has expanded to
\SI{2.19}{\Rsun}. Input files for this and subsequent
\mesa\ calculations are available through Zenodo at
\dataset[10.5281/zenodo.7489814]{\doi{10.5281/zenodo.7489814}}. The
growth of the core is followed using the convective premixing
algorithm \citep{Paxton:2019}, but rotation is neglected. We avoid any
smoothing of the \BV\ frequency profile, as this can introduce small
but consequential departures from mass conservation. Fundamental
parameters for the model are summarized in Table~\ref{t:models}, and
its propagation diagram is plotted in Fig.~\ref{f:prop-diag-koi54}.

\begin{figure*}
  \includegraphics{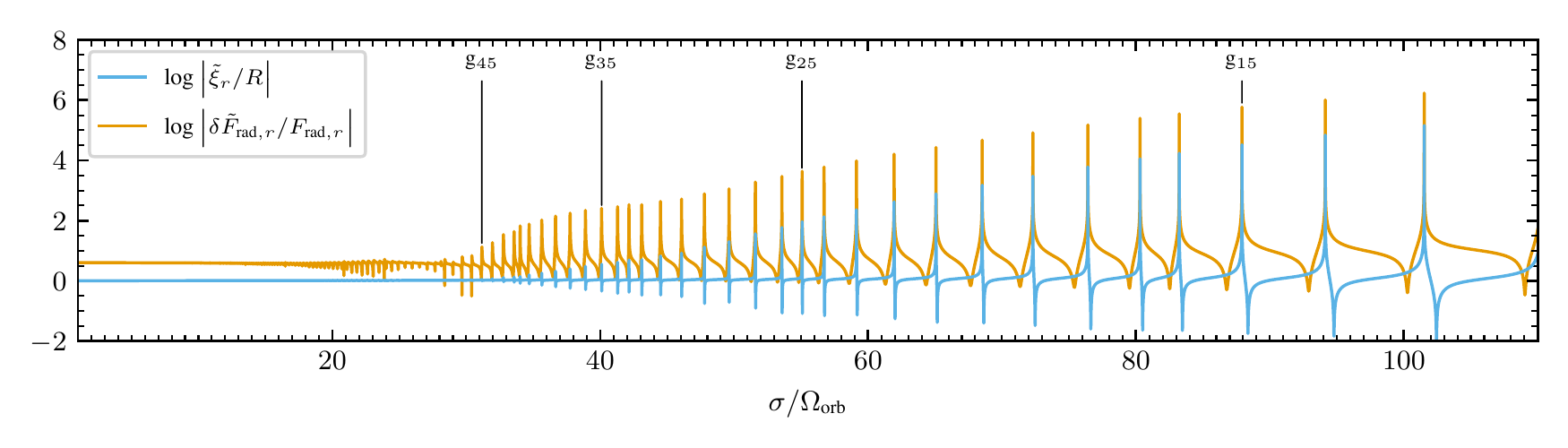}
  \caption{Plot of the radial displacement perturbation $\txir$ and
    the Lagrangian radial flux perturbation $\delta \tFradr$ at the
    surface of the KOI-54 model, as a function of forcing frequency
    $\sigma$. Selected peaks are labeled by the resonant mode's
    classification within the Eckart-Osacki-Scuflaire scheme.}
  \label{f:scan-sigma}
\end{figure*}

Figure 6 of B12 plots the modulus of $\txir$ and $\delta \tFradr$ as a
function of $\sigma/\Oorb$, for a single partial potential with
$\ell=2$. The choices of $m$ and $k$ are left undetermined because B12
neglect rotation in their figure, and treat $\sigma$ as a free
parameter rather than being constrained by
equation~(\ref{e:sigmk}). They normalize the strength of the partial
potential in a manner equivalent to setting $y_{\rm T} = 1$ in
equation~(\ref{e:inhom-bc}). We use \xgyretides\ to repeat these steps
for our KOI-54 model, plotting the results in Fig.~\ref{f:scan-sigma}.

This figure shows qualitative agreement with Fig.~6 of B12. For
$\sigma/\Oorb \gtrsim 30$ the surface perturbations exhibit distinct
peaks, corresponding to resonances with the star's $\ell=2$
free-oscillation modes. Selected peaks are labeled with the resonant
mode's classification within the Eckart-Osaki-Scuflaire scheme
\citep[e.g.,][]{Unno:1989}. For $\sigma/\Oorb \lesssim 30$ the peaks
merge together and dissolve, because the periods of the resonant modes
become appreciably shorter than the local thermal timescale
\begin{equation} \label{e:t-thm}
  \tthm(r) \equiv \int_{r}^{R} \frac{4 \pi r^{2} \rho T \cP}{\Lpri} \diff{r}
\end{equation}
in the outer parts of the main mode-trapping cavity (see
Fig.~\ref{f:prop-diag-koi54}), resulting in significant
radiative-diffusion damping that broadens and suppresses the
resonances. Eventually, for $\sigma/\Oorb \lesssim 15$ the surface
perturbations reach the limits $\txir/\Rpri \rightarrow 1$ and $\delta
\tFradr/\Fradr \rightarrow 4$ corresponding to the $\ell=2$
equilibrium tide (see Section~6.2 of B12 for a discussion of these
limits).

On closer inspection, some differences between the two figures are
apparent. While the KOI-54 stellar models in B12 and the present work
are similar (in particular, sharing the same $\Mpri$ and $\Rpri$),
they are not identical; therefore, the locations and heights of the
resonance peaks are not the same in each figure. More significantly,
over the range $5 \lesssim \sigma/\Oorb \lesssim 30$ the flux
perturbation behaves much more smoothly in Fig.~\ref{f:scan-sigma}
than in Fig.~6 of B12. The reason for this discrepancy is not obvious,
but we speculate that it may be linked to differences in the
near-surface convection of the models. The propagation diagram for our
KOI-54 model (Fig.~\ref{f:prop-diag-koi54}) shows a He\textsc{ii}
convection zone at $\log(P/\si{\dyne\per\cm\squared}) \approx 5.5$,
but this zone is absent in B12's model (cf. their Fig.~1).


\subsection{Direct Solution versus Mode Decomposition in a KOI-54 Model} \label{s:calcs-ds-vs-md}

\begin{figure*}
  \includegraphics{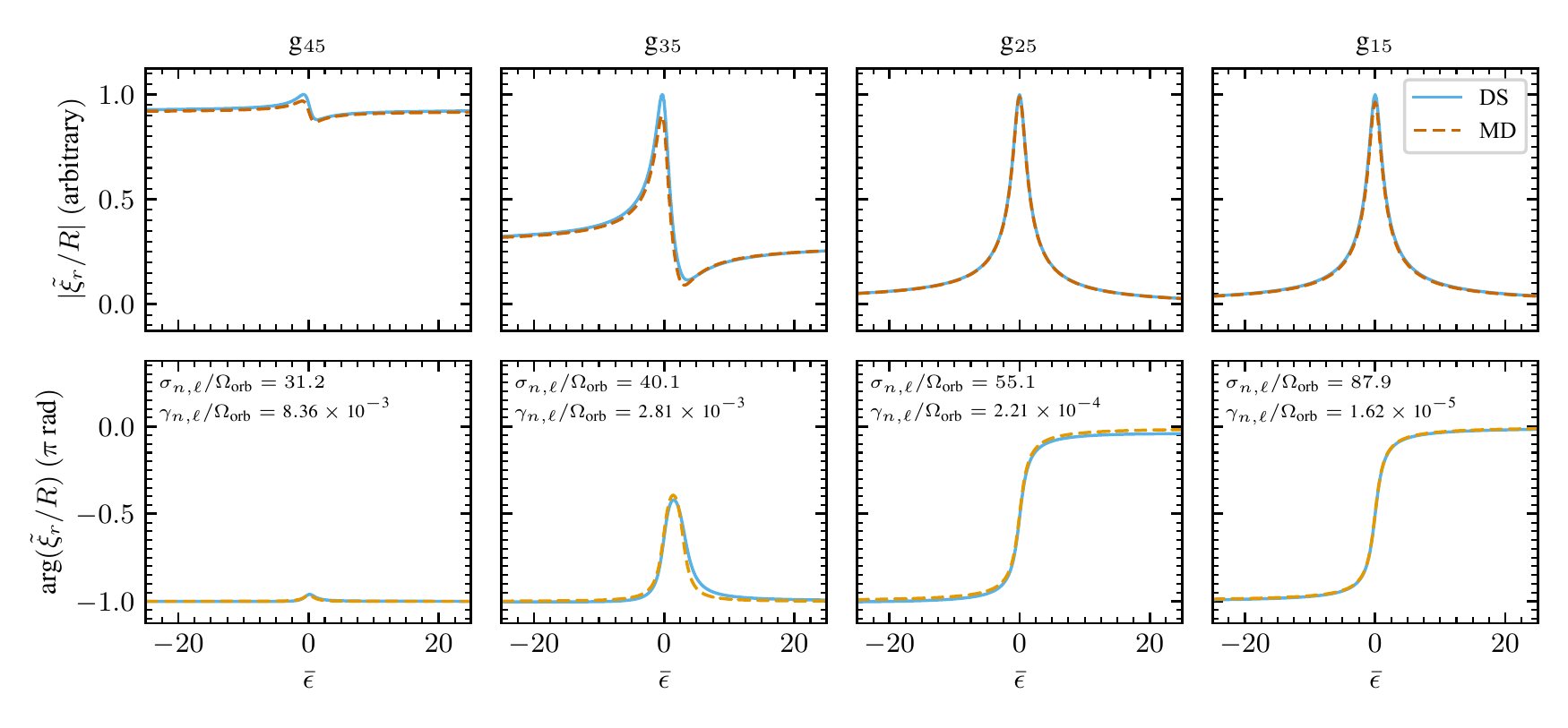}
  \caption{Zoom-in on the four labeled resonance peaks in
    Fig.~\ref{f:scan-sigma}, plotting the complex amplitude (upper
    panels; rescaled to have a maximum value of unity) and phase
    (lower panels) of $\txir/R$ as a function of normalized detuning
    parameter $\detune$ (equation~\ref{e:detune}). Separate curves are
    shown for the DS (\xgyretides) and MD approaches.}
  \label{f:scan-sigma-zoom}
\end{figure*}

\begin{figure}
  \includegraphics{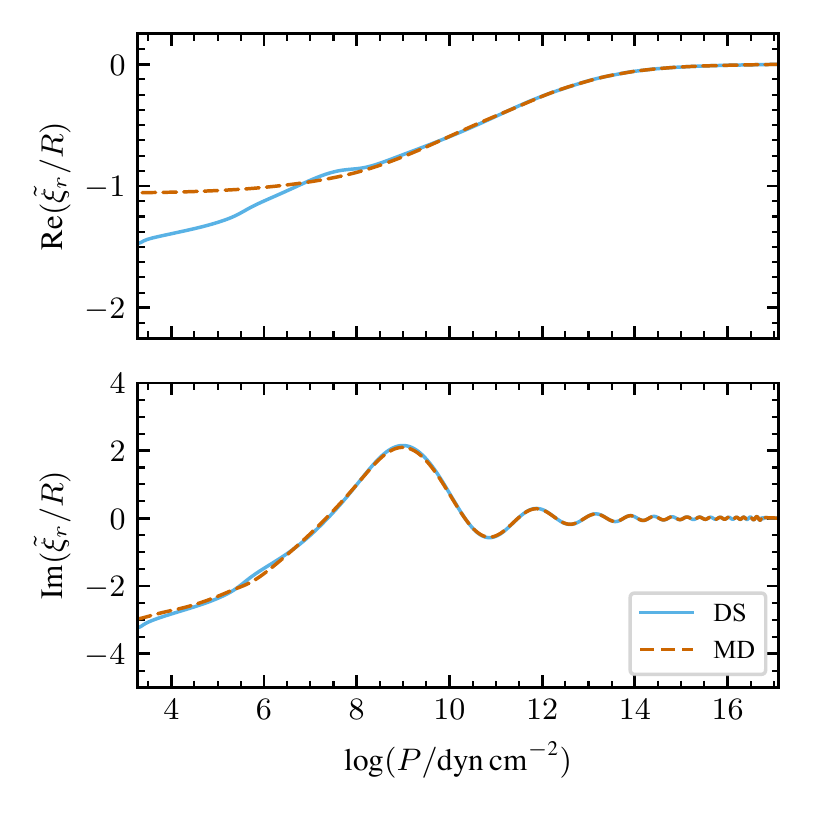}
  \caption{Real (upper panel) and imaginary (lower panel) parts of the
    wavefunctions associated with the radial displacement perturbation
    $\txir$ at the peak of the $\gmode{35}$ resonance, plotted as a
    function of pressure coordinate. The separate curves correspond to
    the DS (\xgyretides) and MD approaches.}
  \label{f:eigfuncs}
\end{figure}

As a validation of the results presented in Fig.~\ref{f:scan-sigma},
we re-calculate the surface perturbations using the MD approach. We
follow the formalism laid out in Section 3.2 of B12, although adopting
non-adiabatic eigenfrequencies $\signl$ and damping rates $\gamnl$
provided by \xgyre\ to evaluate the Lorentzian factor
$\Delta_{n,\ell,m,k}$ (their equation 13; here, $n$ is the mode radial
order). To evaluate the overlap integrals $\overlap$ that weight the
contribution of each free-oscillation mode in the MD superposition, we
use the third expression of equation~(9) in B12; we find that the the
first expression yields unreliable values when $|n| \gtrsim 20$,
because the integrand is highly oscillatory and suffers from
significant cancellation.

Figure~\ref{f:scan-sigma-zoom} zooms in on the four labeled
resonance peaks from Fig.~\ref{f:scan-sigma}, plotting the
complex amplitude and phase of $\txir$ as a function of normalized
detuning parameter
\begin{equation} \label{e:detune}
  \bar{\epsilon} \equiv \frac{\sigma - \signl}{|\gamnl|},
\end{equation}
for the two approaches.  DS (i.e., \xgyretides) and MD agree at higher
forcing frequencies (right-hand panels), but show mismatches toward
lower frequencies (left-hand panels).

To delve further into these discrepancies, Fig.~\ref{f:eigfuncs} plots
the $\txir$ wavefunction evaluated at the peak ($\ndetune=0$) of the
$\gmode{35}$ resonance. The imaginary part of the wavefunction is
spatially oscillatory because it is dominated by the dynamical tide,
comprising the resonantly forced oscillation mode. The real part is
non-oscillatory and corresponds to the equilibrium tide, comprising
the superposition of the other, non-resonant $\ell=2$ modes.

The discrepancies in the wavefunction are restricted to the outer
parts of the stellar envelope. For $\real(\txir)$, the DS and MD
curves begin to diverge at $\log (P/\si{\dyne\per\cm\squared}) \approx
8$ (corresponding to $r/\Rpri \gtrsim 0.96$), while for $\imag(\txir)$
the divergence begins further out at $\log
(P/\si{\dyne\per\cm\squared}) \approx 6$ ($r/\Rpri \approx 0.99$). We
hypothesize that these divergences arise because $\sigma \ll
2\pi/\tthm$ in these superficial layers, leading to significant
non-adiabaticity that MD is unable to correctly reproduce (see Section
6.2 of B12; also, Section 6 of \citealp{Fuller:2017}).


\subsection{Circularization in a B-star Model} \label{s:calcs-circ}

\begin{figure}
  \includegraphics{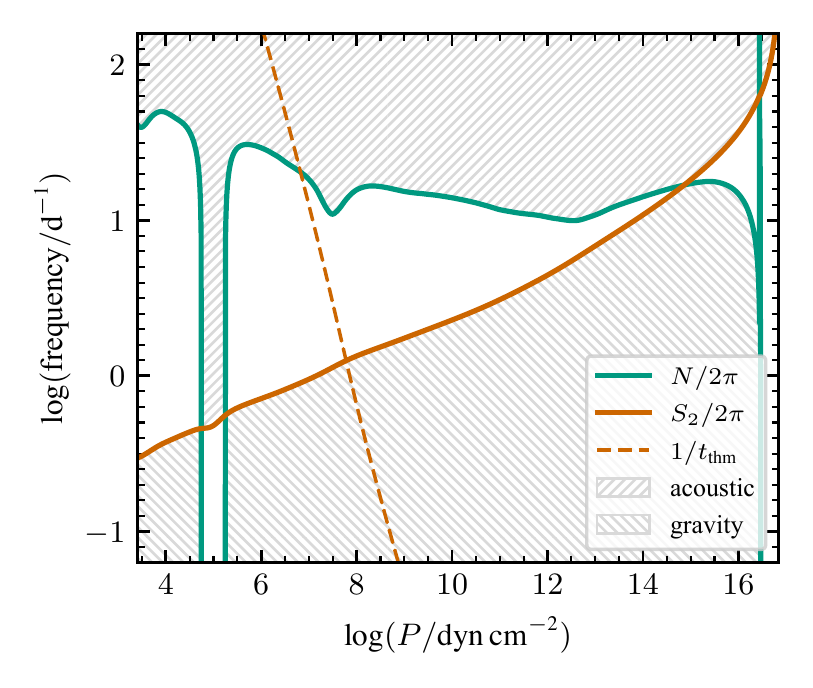}
  \caption{Propagation diagram for the B-star model introduced in the
    text (cf. Fig.~\ref{f:prop-diag-koi54}). The dip in the
    \BV\ frequency around $\log (P/\si{\dyne\per\cm\squared})
    \approx 7.5$ is caused by the iron opacity bump responsible for
    the overstability of the $\gmode{9}-\gmode{17}$ modes.}
  \label{f:prop-diag-5msun}
\end{figure}

\begin{figure*}
  \includegraphics{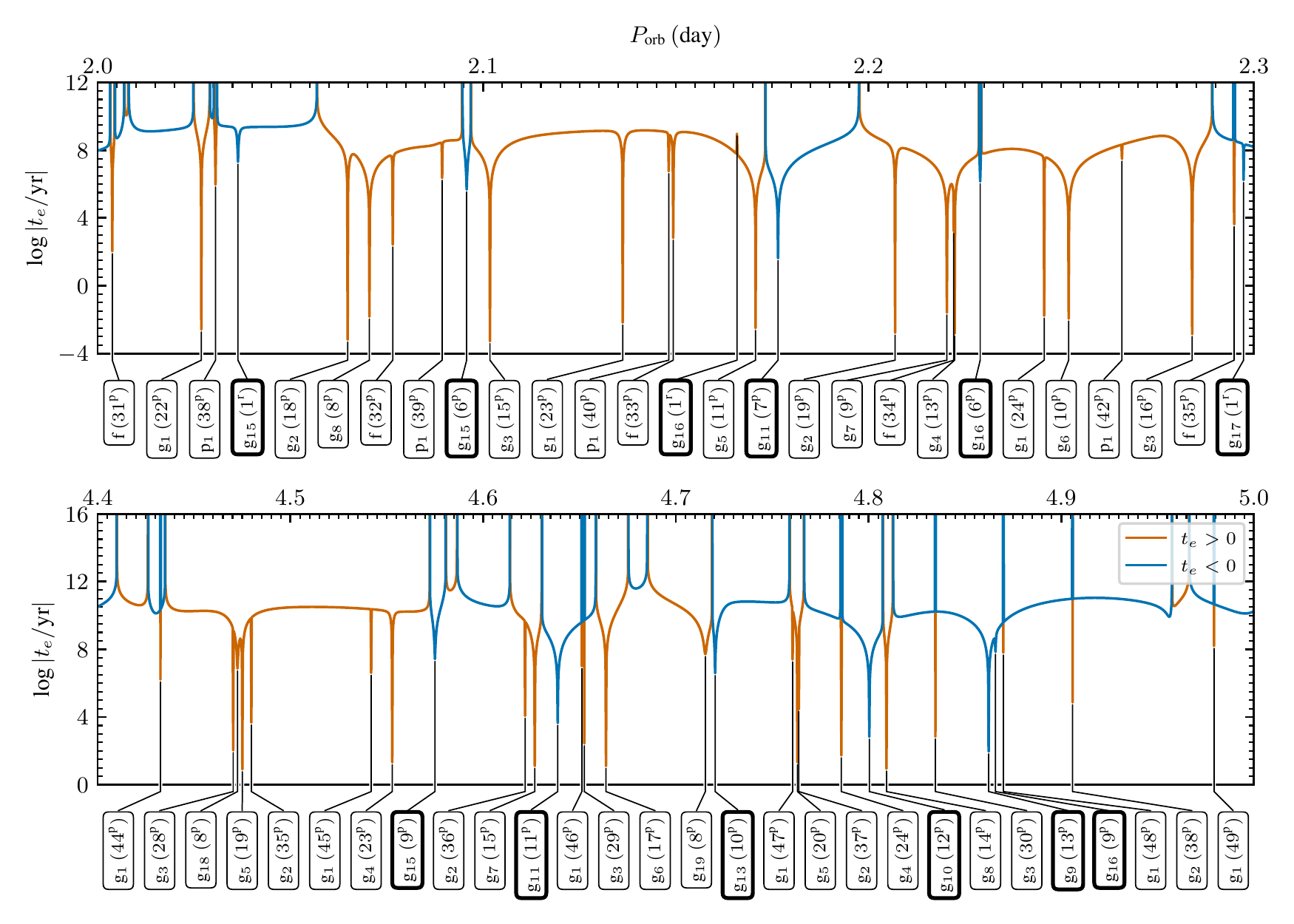}
  \caption{The circularization timescale $\tcirc$ plotted as a
    function of orbital period $\Porb$, for the binary system with the
    B-star primary. Resonances are labeled beneath with the mode
    classification, and in parentheses the harmonic index $k$ and the
    sense of propagation in the co-rotating frame (p=prograde,
    r=retrograde). If the resonance is with an overstable mode, then
    the label border is bolded.}
  \label{f:secular}
\end{figure*}

\begin{figure}
  \includegraphics{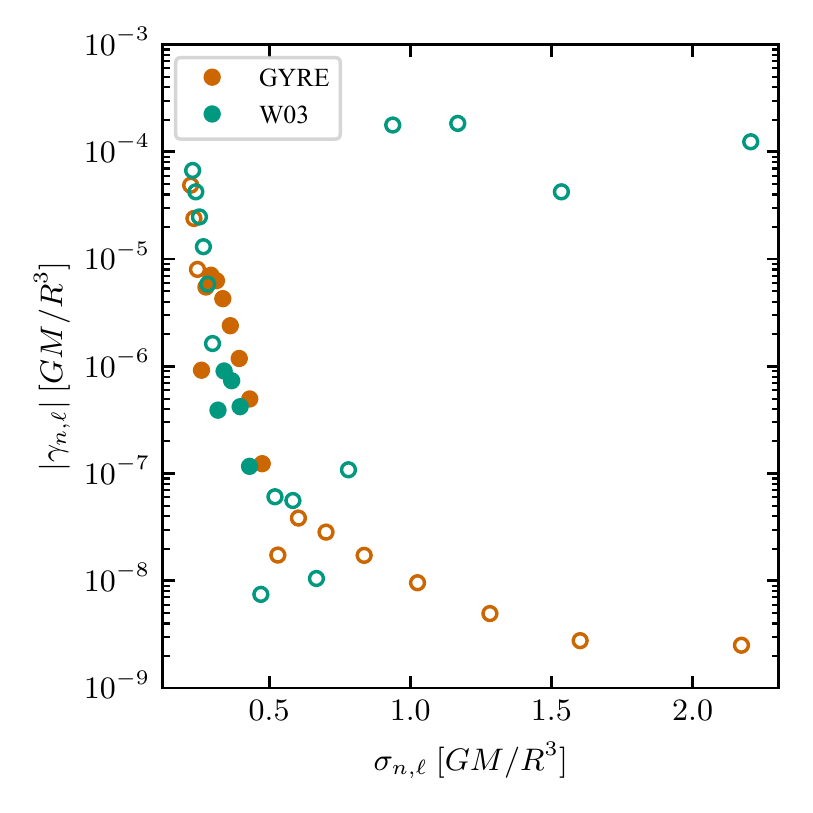}
  \caption{Damping rates $\gamnl$ plotted against
    eigenfrequencies $\signl$ for $\gmode{1}$--$\gmode{20}$ modes of the
    B-star model, as calculated using \xgyre\ and as tabulated
    by W03. Open symbols indicate stable ($\gamnl > 0$) modes, and
    filled symbols overstable ($\gamnl < 0$) modes.}
  \label{f:eigfreqs}
\end{figure}

\citet[][hereafter W03]{Willems:2003} explore the secular orbital
changes due to tides in a binary system comprising a \SI{5}{\Msun}
B-star primary and a \SI{1.4}{\Msun} secondary on an $e=0.5$
orbit. They adopt the MD approach, but include only a single term at a
time in the modal superposition. The calculations illustrated in
Fig.~3 of W03 are used by \citet{Valsecchi:2013} to benchmark their
\cafein\ code, motivating us to do likewise here with \xgyretides.

We use \mesa\ as before to evolve a \SI{5}{\Msun} model from the ZAMS
until its photospheric radius has grown to match the \SI{2.80}{\Rsun}
of W03's model. Fundamental parameters for this model are
summarized in Table~\ref{t:models}, and its propagation diagram is
plotted in Fig.~\ref{f:prop-diag-5msun}. Then, we apply
\xgyretides\ to evaluate the star's response to the dominant
contributions in the tidal potential~(\ref{e:pottide}), comprising the
$\ell=2$ terms with $|m| = 2$ and $m k > 0$. We further restrict the
summation over $k$ to terms with a magnitude at least $10^{-12}$ times
that of the largest term.  As in W03, we assume the stellar angular
rotation frequency is equal to the periastron angular velocity of the
secondary,
\begin{equation}
  \Operi = \Oorb \, \sqrt{\frac{(1+e)}{(1-e)^{3}}}.
\end{equation}
Based on this configuration, Fig.~\ref{f:secular} plots the circularization
timescale
\begin{equation}
  \tcirc \equiv - \left[ \frac{1}{e} \left( \deriv{e}{t} \right) \right]^{-1}_{\rm sec}
\end{equation}
as a function of orbital period $\Porb \equiv 2\pi/\Oorb$ over a pair
of intervals\footnote{These correspond to the short- and long-period
limits in Fig.~3 of W03.}. The secular rate-of-change of eccentricity
is evaluated via
\begin{multline} \label{e:secular-edot}
  \left( \deriv{e}{t} \right)_{\rm sec} = 4 \,
  \Oorb \,
  \mratio
  \sum_{\ell,m,k \geq 0}
  \left( \frac{\Rpri}{a} \right)^{\ell+3} \,
  \left( \frac{\rs}{\Rpri} \right)^{\ell+1} \\
  \mbox{} \times \kap[\ell,m,k] \,
  \imag(\Fbar[\ell,m,k]) \,
  \Gbar[\ell,m,k]{3};
\end{multline}
this comes from equation~(55) of \citet{Willems:2010}, with
$F_{\ell,m,k}$ replaced by $\Fbar[\ell,m,k] \equiv F_{\ell,m,-k}$ and
$G^{(3)}_{\ell,m,k}$ by $\Gbar[\ell,m,k]{3} \equiv G^{(3)}_{\ell,m,-k}$ to account for
the differing sign convention in the assumed time dependence of
partial tides (see equation~\ref{e:angtime}). Note that the summation
is now restricted to $k \geq 0$.

The $|\tcirc|$ data plotted in the figure vary smoothly with $\Porb$
between a series of sharp extrema. The maxima (more correctly,
singularities) arise when $(\sderiv{e}{t})_{\rm sec}$ passes through
zero. The minima arise from resonances with the star's $\ell=2$
free-oscillation modes, similar to the peaks seen in
Fig.~\ref{f:scan-sigma}. However, a key difference here is that the
star is being forced with a superposition~(\ref{e:pottide}) of partial
tidal potentials, rather than a single one as before. The criterion
for resonance $\sigmk \approx \signl$ can be satisfied for many
different values of $k$, leading to multiple resonances with the same
mode. This can be seen in the figure; for instance, the upper panel
shows resonances between the $f$ mode and the partial tidal potentials with
$k=31,\ldots,35$.

In the vicinity of some of the resonances shown in the figure, $\tcirc
< 0$ (blue) indicates that the tide acts to increase the eccentricity
of the orbit, driving it further away from circular. This behavior is
an instance of the `inverse tides' phenomenon discussed by
\citet{Fuller:2021}, and occurs when the summation in
equation~(\ref{e:secular-edot}) is dominated by a single,
\emph{positive} term. There are four distinct configurations that lead
to this outcome:
\begin{enumerate}[label=\Roman*.]
\item $\Gbar[\ell,m,k]{3} > 0$ and $\imag(\Fbar[\ell,m,k]) > 0$, the latter because
  \begin{enumerate}
  \item the resonant mode is prograde in the co-rotating frame ($\signl/m > 0$)
    and stable ($\gamnl > 0$); or \label{l:pos-pro-s}
  \item the resonant mode is retrograde in the co-rotating frame
    ($\signl/m < 0$) and overstable ($\gamnl < 0$). \label{l:pos-ret-o}
  \end{enumerate}
\item $\Gbar[\ell,m,k]{3} < 0$ and $\imag(\Fbar[\ell,m,k]) < 0$, the latter because
  \begin{enumerate}
  \item the resonant mode is prograde in the co-rotating frame ($\signl/m > 0$)
    and overstable ($\gamnl < 0$); or \label{l:neg-pro-o}
  \item the resonant mode is retrograde in the co-rotating frame
    ($\signl/m < 0$) and stable ($\gamnl > 0$). \label{l:neg-ret-s}

  \end{enumerate}
\end{enumerate}
All of the $\tcirc < 0$ resonances seen in Fig.~\ref{f:secular} are
instances of cases~(\ref{l:pos-ret-o}) or~(\ref{l:neg-pro-o}), and
therefore involve overstable modes. The overstability is caused by the
iron-bump opacity mechanism responsible for the slowly pulsating B
(SPB) stars \citep[e.g.,][]{Gautschy:1993,Dziembowski:1993}; in the
B-star model considered here, which falls well inside the SPB
instability strip \citep[e.g.,][]{Pamyatnykh:1999,Paxton:2015}, this
mechanism excites the $\ell=2$ $\gmode{9}$--$\gmode{17}$ modes.

Comparing Fig.~\ref{f:secular} against Fig.~3 of W03 reveals some
important differences. The latter shows numerous gaps and
discontinuities in $\tcirc$, that appear to arise because W03 only
allow a given mode to contribute toward the MD superposition when its
detuning parameter (equation~\ref{e:detune}) satisfies $0.1\,\epstide
\leq |\ndetune \gamnl/ \signl| \leq 10\,\epstide$. The lower bound on
$\ndetune$ means that the central parts of each resonance are omitted,
and so Fig.~3 of W03 does not fully reveal how small $\tcirc$ can
become when close to a resonance.

At the short-period limit of the range shown in the figures, there is
also disagreement between the typical magnitude of $\tcirc$ between
the resonances; Fig.~\ref{f:secular} shows an inter-resonance
$|\tcirc| \approx \SI{e9}{\year}$ , whereas for W03 it is 2--4 orders
of magnitude shorter. This is likely a consequence of W03
over-estimating the damping for the $\gmode{1}$--$\gmode{4}$ modes, which
dominate the tidal response at short orbital
periods. Figure~\ref{f:eigfreqs} plots eigenfrequencies $\signl$ and
damping rates $\gamnl$ for the $\gmode{1}$--$\gmode{20}$ modes, as calculated
using \xgyre\ and as obtained from Table~1 of W03. The data for the
$\gmode{5}$--$\gmode{16}$ modes are in reasonable agreement, especially given
that the underlying stellar models do not have the exact same internal
structure. However, the W03 damping rates for the $\gmode{1}$--$\gmode{4}$ modes
are four-to-five orders of magnitude larger than the \xgyre\ ones.


\subsection{Pseudo-Synchronization in a KOI-54 Model} \label{s:calcs-sync}

\begin{figure}
  \includegraphics{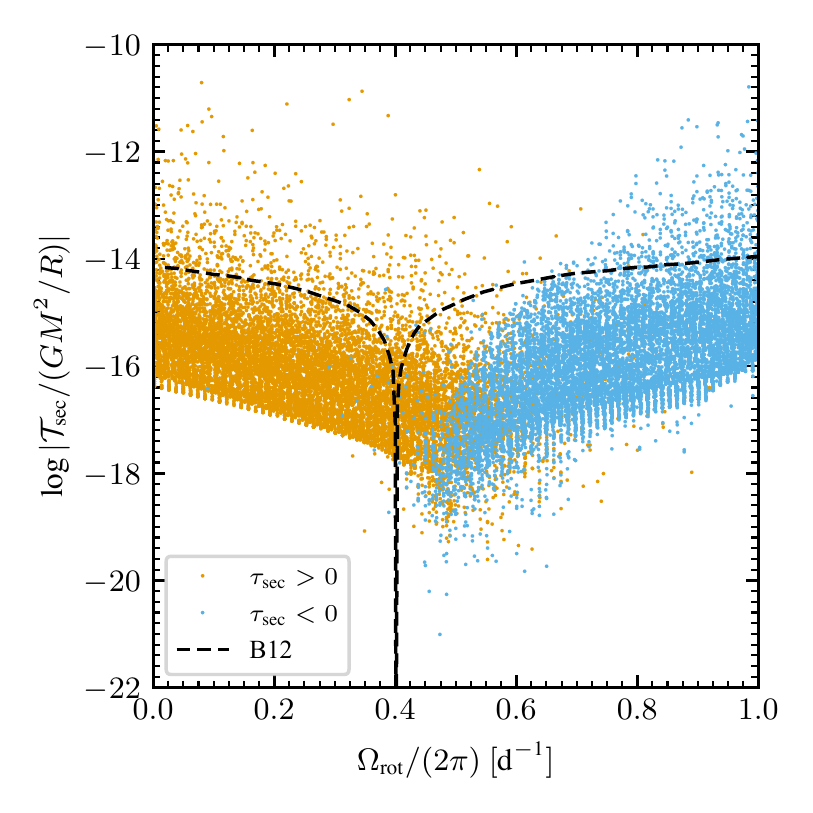}
  \caption{Secular torque $\torsec$ plotted against stellar
    angular rotation frequency $\Orot$ for the KOI-54 primary
    model. The dashed line shows the equilibrium-tide torque extracted
    from Fig. 4 of B12.}
  \label{f:torque}
\end{figure}

B12 explore how tides can modify the rotation of the primary star in
the KOI-54 system, by evaluating the secular tidal torque $\torsec$ on
the primary star as a function of the star's rotation rate. Using an
MD approach (see their Appendix C), they consider contributions toward
the torque from the $\ell=2$ partial tidal potentials. Their Fig. 4
shows a smoothly varying $\torsec$ punctuated by many narrow peaks due
to modal resonances. The smooth part corresponds to the torque from
the equilibrium tide, and passes through zero at the
pseudo-synchronous angular frequency
\begin{equation}
  \Opsy \equiv \Oorb \frac{1 + (15/2)e^{2} + (45/8)e^{4} + (5/16)e^{6}}{[1 + 3e^{2} + (3/8)e^4](1 - e)^{3/2}}
\end{equation}
first derived by \citet{Hut:1981} in his theory of tides in the weak
friction limit. For the orbital parameters of KOI-54, the
pseudo-synchronous frequency is $\Opsy/(2\pi) = \SI{0.395}{\per\day}$.

To repeat this calculation, we use \xgyretides\ to evaluate the
response of the KOI-54 model (Section~\ref{s:calcs-surf}) to the
$\ell=2$ terms in the tidal potential~(Equation~\ref{e:pottide}) for
\num{25000} rotation frequencies in the interval $\SI{0}{\per\day}
\leq \Orot/(2\pi) \leq \SI{1}{\per\day}$. As in the preceding section,
we restrict the summation over $k$ to terms with a magnitude at least
$10^{-12}$ times that of the largest term. Then, we evaluate the tidal
torque via
\begin{multline}
  \torsec = 
  4 \Oorb \sqrt{\frac{G \Mpri^{3} \mratio^{2}}{1 + \mratio}} \, \mratio \, a^{1/2}
  \sum_{\ell,m,k \geq 0}
  \left( \frac{\Rpri}{a} \right)^{\ell+3} \,
  \left( \frac{\rs}{\Rpri} \right)^{\ell+1} \\
  \times \kap[\ell,m,k] \,
  \imag(\Fbar[\ell,m,k]) \,
  \Gbar[\ell,m,k]{4};
\end{multline}
this comes from equation~(63) of \citet{Willems:2010}, with
$G^{(4)}_{\ell,m,k}$ replaced by $\Gbar[\ell,m,k]{4} \equiv
G^{(4)}_{\ell,m,-k}$.

Fig.~\ref{f:torque} plots $\torsec$ as a function of $\Orot$, using
discrete points rather than a continuous line because we are
significantly undersampling the dense forest of resonances. Also
plotted for comparison is the smooth (equilibrium tide) part of the
torque extracted from Fig.~4 of B12.  Clearly, there are some
significant differences between the two figures. Ours shows a torque
that's generally positive torque at small rotation frequencies, and
negative at high frequencies; however, the switch-over point is not
nearly as sharply defined as in B12, and occurs at a frequency
$\Orot/(2\pi) \approx \SI{0.5}{\per\day}$ around 25\% higher than
$\Opsy$. Moreover, on either side of the switch-over, the lower
envelope of our torque values is around two orders of magnitude
smaller than the B12 curve.

Exploratory calculations indicate that these differences are not a
result of inaccurate overlap integrals (as was the case in
Section~\ref{s:calcs-ds-vs-md}), but rather due to a genuine
incompatibility between the DS and MD approaches. While a detailed
investigation of this problem is beyond the scope of the present
paper, we believe the fault lies with MD's over-estimation of
radiative dissipation for the equilibrium tide. If this hypothesis is correct, an immediate corollary is that pseudo-synchronization as envisaged by \citet{Hut:1981} does not operate for stars in which radiative dissipation dominates the tidal damping.


\section{Summary and Discussion} \label{s:discuss}

To briefly summarize the preceding sections: we establish the
theoretical foundations for our tides treatment
(Section~\ref{s:formalism}), describe modifications to \gyre\ to
implement tides (Section~\ref{s:implement}), and then apply the new
\xgyretides\ executable to selected problems (Section~\ref{s:calcs}).
These example calculations uncover disagreements between the DS and MS
approaches, arising for a variety of reasons --- from numerical
inaccuracies in overlap integrals (Section~\ref{s:calcs-ds-vs-md}),
through to what appears to be an incompatibility between the
approaches. We look forward to future opportunities to investigate
these disagreements.

We also plan a number of enhancements to \xgyretides\ that will extend
its capabilities. Key milestones include adding the ability to model
systems with spin-orbit misalignments \citep[e.g., following the
  formalism described by][]{Fuller:2017}; the incorporation of
additional damping mechanisms beyond radiative diffusion \citep[for
  instance, turbulent viscosity within convection zones; see
][]{Willems:2010}; and treating the effects of the Coriolis force,
which was neglected in deriving the linearized equations
(Appendix~\ref{a:lin-eqns}). Partial treatment of the Coriolis force
is already included in the main \xgyre\ executable, via the
traditional approximation of rotation (TAR; see, e.g.,
\citealp{Bildsten:1996,Lee:1997,Townsend:2003}). However, in its
current form this implementation cannot be used in \xgyretides,
because the angular operator appearing in the linearized continuity
equation within the TAR \citep[see equation~6 of][]{Bildsten:1996}
does not commute with the angular part of the Laplacian operator
appearing in the linearized Poisson equation~(\ref{e:lin-poisson}).

\gyre\ is not the first software package that adopts the DS approach
to model stellar tides; \citet{Pfahl:2008} and \citet{Valsecchi:2013}
describe functionally similar codes. Although the former authors' code
has never been made publicly available, the latters' \cafein\ code is
accessible on
GitHub\footnote{https://github.com/FrancescaV/CAFein}. After fixing a
number of bugs in \cafein\ (for instance, relating to misinterpreting
cell-centered quantities in \mesa\ models as face-centered), we have
undertaken exploratory calculations comparing it against \xgyretides,
and find the two codes are in general agreement. Given that
\cafein\ is unmaintained, we decided that more-detailed comparison
would not be a worthwhile exercise.

It is our hope that \xgyretides\ will provide a standardized and
well-supported community tool for simulating tides of spherical stars
within the linear limit. Specific areas where we anticipate productive
applications include modeling the many heartbeat systems discovered by
Kepler \citep[e.g.,][]{Thompson:2012}, TESS
\citep[e.g.,][]{Kolaczek:2021} and OGLE \citep{Wrona:2022};
investigating why most of these systems rotate faster than the
pseudo-synchronous rate \citep{Zimmerman:2017}; and exploring orbital
and rotational evolution in more-general star-star and star-planet
systems. These latter activities will initially be restricted to cases
where radiative diffusion dominates the tidal damping \citep[e.g., the
  $\gamma$ Doradus stars considered by ][]{LiGang:2020}; but with the
planned addition of other damping mechanisms, they can be extended
more broadly.


\section*{Acknowledgments}

This work has been supported by NSF grants ACI-1663696, AST-1716436
and PHY-1748958, and NASA grant 80NSSC20K0515. This research was also supported by STFC through grant ST/T00049X/1. The authors thank the referee for comments that have improved this paper.

\facilities{We have made extensive use of NASA's
  Astrophysics Data System Bibliographic Services.}

\software{Astropy \citep{astropy:2013,astropy:2018,astropy:2022},
  \gyre\ \citep{Townsend:2013,Townsend:2018,Goldstein:2020},
  Matplotlib \citep{Hunter:2007}, \mesa\ 
  \citep{Paxton:2011,Paxton:2013,Paxton:2015,Paxton:2018,Paxton:2019,Jermyn:2022}}


\bibliography{gyre-tides}


\appendix

\section{Expansion Coefficients} \label{a:exp-coeff}

The expansion coefficients appearing in
equation.~(\ref{e:pottide-part}) are given by
\begin{equation}
  \cbar[\ell,m,k] = \frac{4\pi}{2\ell+1} \,
  \left( \frac{\Rpri}{a} \right)^{\ell-2} \,
       {Y^{m}_{\ell}}^{*}(\pi/2,0) \,
       X^{-(\ell+1),-m}_{-k},
\end{equation}
where $Y^{m}_{\ell}$ is a spherical harmonic
(Appendix~\ref{a:sph-harm}) and $X^{-(\ell+1),-m}_{-k}$ is a Hansen
coefficient (Appendix~\ref{a:hansen-coeff}). They are related to the
$c_{\ell,m,k}$ coefficients defined by \citet{Willems:2010} via
\begin{equation}
  \cbar[\ell,m,k] =
  (-1)^{(|m|-m)/2} \,
  \sqrt{\frac{4\pi}{2\ell+1}\frac{(l+|m|)!}{(l-|m|)!}} \,
  c_{\ell,m,-k}.
\end{equation}

\section{Spherical Harmonics} \label{a:sph-harm}

There are a number of alternate definitions of the spherical
harmonics, differing in normalization and phase conventions. We follow
\citet{Arfken:2013} and adopt
\begin{equation} \label{e:sph-harm}
  Y^{m}_{\ell}(\vartheta,\varphi) = \sqrt{\frac{2\ell+1}{4\pi} \frac{(\ell-m)!}{(\ell+m)!}} P^{m}_{\ell}(\cos\vartheta) \ee{i m \varphi}.
\end{equation}
The associated Legendre functions are in turn defined by the Rodrigues
formula
\begin{equation} \label{e:assoc-leg}
  P^{m}_{\ell}(x) = \frac{(-1)^{m}}{2^{\ell} \ell!} \left( 1 - x^{2} \right)^{m/2} \deriv[\ell+m]{}{x} \left( x^{2} - 1 \right)^{\ell}
\end{equation}
[the extra $(-1)^{m}$ factor is the Condon-Shortley phase term]. With
these definitions, the spherical harmonics obey the orthonormality
condition
\begin{equation}
  \int_{0}^{2\pi} \int_{0}^{\pi} Y^{m}_{\ell} \, Y^{m'*}_{\ell'} \, \sin\vartheta \, \diff{\vartheta} \diff{\varphi} = \delta _{\ell,\ell'} \delta_{m,m'},
\end{equation}
and moreover the relation
\begin{equation}
  {Y^{m}_{\ell}}^{*} = (-1)^{m} Y^{-m}_{\ell}.
\end{equation}
%


\section{Hansen Coefficients} \label{a:hansen-coeff}

The Hansen coefficients are defined implicitly by the equation
\begin{equation}
  \left( \frac{\rsec}{a} \right)^{n} \ee{\ii m \tanom} = \sum_{k=-\infty}^{\infty} X^{n,m}_{k} \ee{\ii k \manom}
\end{equation}
\citep[e.g.,][]{Hughes:1981}. They can be evaluated via
\begin{equation}
  X^{n,m}_{k} = \frac{(1 - e^{2})^{n}}{2\pi} \int_{-\pi}^{\pi} (1 + e \cos\manom)^{-n} \cos(m\tanom - k \manom) \,\diff{\manom};
\end{equation}
however, an equivalent form due to \citet{Smeyers:1991},
\begin{equation}
  X^{n,m}_{k} = \frac{(1 - e^{2})^{n+3/2}}{2\pi} \int_{-\pi}^{\pi} (1 + e \cos\manom)^{-n-2} \cos(m\tanom - k \manom) \,\diff{\tanom},
\end{equation}
is more convenient because it does not require Kepler's
equation~(\ref{e:kepler}) be solved for $\eanom$. The integrand is
periodic with respect to $\tanom$, and so the exponential convergence
of the trapezoidal quadrature rule \citep{Trefethen:2014} is ideal for
evaluating this integral.


\section{Linearized Equations} \label{a:lin-eqns}

We introduce the tidal potential $\pottide$ (equation~\ref{e:pottide})
into the fluid equations governing the primary star as a small
($\epstide \ll 1$) perturbation about the equilibrium state. We assume
this equilibrium state is spherically symmetric and static; while we
allow for uniform rotation about the $z$-axis with angular velocity
$\Orot$, we neglect the inertial (Coriolis and centrifugal) forces
arising from this rotation. Discarding terms of second- or
higher-order in $\epstide$ from the perturbed structure equations, and subtracting away the equilibrium state, yields linearized versions of the fluid equations. These comprise the mass equation
\begin{equation} \label{e:lin-mass}
  \left( \pderiv{}{t} + \Orot \pderiv{}{\varphi} \right) \rho' + \nabla \cdot ( \rho \vv' ) = 0;
\end{equation}
the momentum equation
\begin{equation} \label{e:lin-mom}
  \sum_{i=\{r,\vartheta,\varphi\}}
  \left[ \left( \pderiv{}{t} + \Orot \pderiv{}{\varphi} \right) v'_{i} \right] \vei =
  - \frac{1}{\rho} \nabla P' + \frac{\rho'}{\rho^{2}} \deriv{P}{r} - \nabla \pottot',
\end{equation}
where $\pottot' \equiv \potpri' + \pottide$; Poisson's equation
\begin{equation} \label{e:lin-poisson}
  \nabla^{2} \pottot' = 4 \pi G \rho';
\end{equation}
the heat equation
\begin{equation} \label{e:lin-heat}
  T \left( \pderiv{}{t} + \Orot \pderiv{}{\varphi} \right) \delta S = \delta \epsnuc - \delta \left[ \frac{1}{\rho} \nabla \cdot (\vFrad + \vFcon) \right];
\end{equation}
and the radiative diffusion equation
\begin{equation} \label{e:lin-rad-diff}
\delta \vFrad = \left( 4 \frac{\delta T}{T} - \frac{\delta \rho}{\rho} - \frac{\delta \kappa}{\kappa} \right) \vFrad + 
\frac{\delta \left( \nabla \ln T \right)}{\sderiv{\ln T}{r}} \Fradr.
\end{equation}
In these equations, $\vv$ is the fluid velocity; $P$, $T$, $\rho$, and
$S$ are the pressure, density, temperature, and specific entropy,
respectively; $\vFrad$ and $\vFcon$ are the radiative and convective
flux vectors, with $\Fradr$ the radial component of the former;
$\kappa$ is the opacity and $\epsnuc$ the specific nuclear energy
generation rate; and $\potpri$ is the self-gravitational potential. A prime ($'$) suffix on a quantity indicates the Eulerian (fixed position) perturbation, while a $\delta$ prefix indicates the Lagrangian (fixed mass element) perturbation; the absence of either modifier signifies the equilibrium state. To first order, Eulerian and Lagrangian perturbations to a quantity $f$ are linked through
\begin{equation} \label{e:lag-eul}
  \delta f = f' + (\vxi \cdot \nabla) f,
\end{equation}
where the displacement perturbation vector $\vxi$ is related to the velocity
perturbation $\vv'$ via
\begin{equation}
  \vv' = \sum_{i=\{r,\vartheta,\varphi\}} \left[ \left( \pderiv{}{t} + \Orot \pderiv{}{\varphi} \right) \xi_{i} \right] \vei.
\end{equation}

The system of differential
equations~(\ref{e:lin-mass}--\ref{e:lin-rad-diff}) is augmented by a
convective freezing prescription
\begin{equation} \label{e:lin-conv}
  \delta \left( \frac{1}{\rho} \nabla \cdot \vFcon \right) = 0
\end{equation}
\citep[this corresponds to approach 1 in the classification scheme
  by][]{Pesnell:1990}, together with the linearized thermodynamic
  relations
\begin{equation} \label{e:lin-thermo}
  \frac{\delta \rho}{\rho} = \frac{1}{\Gamma_{1}} \frac{\delta P}{P} - \upsT \frac{\delta S}{\cP}, \qquad
  \frac{\delta T}{T} = \nablaad \frac{\delta P}{P} + \frac{\delta S}{\cP},
\end{equation}
and the linearized microphysics equations
\begin{equation} \label{e:lin-micro}
  \frac{\delta \kappa}{\kappa} = \kapad \frac{\delta P}{P} + \kapS \frac{\delta S}{\cP}, \qquad
  \frac{\delta \epsnuc}{\epsnuc} = \epsad \frac{\delta P}{P} + \epsS \frac{\delta S}{\cP}.
\end{equation}
Here,
\begin{equation}
\begin{gathered}
  \Gamma_{1} \equiv \left( \pderiv{\ln P}{\ln \rho} \right)_{S}, \qquad
  \nablaad \equiv \left( \pderiv{\ln T}{\ln P} \right)_{S}, \qquad
  \upsT \equiv -\left( \pderiv{\ln \rho}{\ln T} \right)_{P}, \qquad
  \cP \equiv \left( \pderiv{S}{\ln T} \right)_{P} \\
  \kapad \equiv \left( \pderiv{\ln \kappa}{\ln P} \right)_{S}, \qquad
  \kapS \equiv \cP \left( \pderiv{\ln \kappa}{S} \right)_{P}, \qquad
  \epsad \equiv \left( \pderiv{\ln \epsnuc}{\ln P} \right)_{S}, \qquad
  \epsS \equiv \cP \left( \pderiv{\ln \epsnuc}{S} \right)_{P}.
\end{gathered}
\end{equation}

The system of equations is closed by applying boundary conditions at
the center and surface of the primary star. At the center we require
that perturbations remain regular. At the surface, the boundary
conditions are composed of the vacuum condition
\begin{equation}
  \delta P = 0,
\end{equation}
the linearized Stefan-Boltzmann law
\begin{equation}
  \frac{\delta \Lrad}{\Lrad} = 2 \frac{\xir}{\Rpri} + 4 \frac{\delta T}{T},
\end{equation}
where $\Lrad \equiv 4 \pi r^{2} \Fradr$ is the radiative luminosity,
and the requirement that $\delta \pottot$ and its gradient are
continuous across the surface.


\section{Tidal Equations} \label{a:tidal-eqns}

The tidal equations govern the radial functions appearing in the
solution forms~(\ref{e:sol-xi},\ref{e:sol-f}). To obtain these
equations for a given combination of indices $\{\ell',m',k'\}$, we
substitute these solution forms into the linearized equations
(Appendix~\ref{a:lin-eqns}), multiply by a weighting factor
$\angtime[\ell',m',k']^{*}$, and then integrate over $4\pi$ steradians
and one orbital period. Following these steps, the mass equation~(\ref{e:lin-mass}) becomes
\begin{equation} \label{e:tidal-mass}
  \frac{\delta \trho}{\rho} +
  \frac{1}{r^{2}} \deriv{}{r} \left( r^{2} \txir \right)
  - \frac{\ell(\ell+1)}{r} \txih = 0
\end{equation}
(for notational compactness and clarity, here and subsequently we omit
the $\ell,m,k$ subscripts on dependent variables such as
$\trho[\ell,m,k]$ and $\txir[\ell,m,k]$). The radial and horizontal
components of the momentum equation~(\ref{e:lin-mom}) become
\begin{align} \label{e:tidal-mom}
  \sigmk^{2} \, \txir &= \frac{1}{\rho} \deriv{\tP'}{r} - \frac{\trho'}{\rho^{2}} \deriv{P}{r} + \deriv{\tpottot'}{r}, \\
  \sigmk^{2} \, \txih &= \frac{1}{r} \left( \frac{\tP'}{\rho} + \tpottot' \right),
\end{align}
respectively, where
\begin{equation} \label{e:sigmk}
  \sigmk \equiv k \Oorb - m \Orot
\end{equation}
represents the tidal forcing frequency measured in a frame rotating
with the primary star. Poisson's equation~(\ref{e:lin-poisson}) becomes
\begin{equation} \label{e:tidal-poisson}
  \frac{1}{r^{2}} \deriv{}{r} \left( r^{2} \deriv{\tpottot'}{r} \right) - \frac{\ell(\ell+1)}{r^{2}} \tpottot' = 4 \pi G \trho',
\end{equation}
while the heat equation~(\ref{e:lin-heat}), expressed in terms of $\Lrad$ and its perturbation, becomes
\begin{equation} \label{e:tidal-heat}
- \ii \sigmk \, T \delta \tS =
\delta \teps -
\frac{1}{4\pi r^{2} \rho} \deriv{\delta \tLrad}{r} +
\frac{\ell(\ell+1)}{\sderiv{\ln T}{\ln r}} \frac{\Lrad}{4\pi r^{3}\rho} \frac{\tT'}{T} +
\ell(\ell+1) \frac{\txih}{4\pi r^{3}\rho} \deriv{\Lrad}{r}.
\end{equation}
(equation~\ref{e:lin-conv} has been used to eliminate the
convective terms from this equation.)  The radiative diffusion
equation~(\ref{e:lin-rad-diff}) becomes
\begin{equation} \label{e:tidal-diff}
  \frac{\delta \tLrad}{\Lrad} = - \frac{\delta \tkap}{\kappa} + 4 \frac{\txir}{r} - \ell(\ell+1) \frac{\txih}{r} + 4 \frac{\delta \tT}{T} +
  \frac{1}{\sderiv{\ln T}{\ln r}} \deriv{(\delta \tT/T)}{r}.
\end{equation}
The thermodynamic relations~(\ref{e:lin-thermo}) become
\begin{equation}
  \frac{\delta \trho}{\rho} = \frac{1}{\Gamma_{1}} \frac{\delta \tP}{P} - \upsT \frac{\delta \tS}{\cP}, \qquad
  \frac{\delta \tT}{T} = \nablaad \frac{\delta \tP}{P} + \frac{\delta \tS}{\cP},
\end{equation}
and the microphysics relations~(\ref{e:lin-micro}) become
\begin{equation}
  \frac{\delta \tkap}{\kappa} = \kapad \frac{\delta \tP}{P} + \kapS \frac{\delta \tS}{\cP}, \qquad
  \frac{\delta \teps}{\epsnuc} = \epsad \frac{\delta \tP}{P} + \epsS \frac{\delta \tS}{\cP}.
\end{equation}
The inner boundary conditions are
\begin{equation}
  \txir - \ell\, \txih = 0, \qquad
  \deriv{\tpottot'}{r} - \ell \frac{\tpottot'}{r} = 0, \qquad
  \delta \tS = 0
\end{equation}
evaluated at the center of the star $r=0$. Finally, the outer boundary
conditions are
\begin{equation}
  \delta \tP = 0, \qquad
  \frac{\delta \tLrad}{\Lrad} = 2 \frac{\txir}{\Rpri} + 4 \frac{\delta \tT}{T}, \qquad
  \deriv{\tpottot'}{r} + \frac{\ell+1}{r} \tpottot' + 4 \pi G \rho\, \txir =
  \frac{2 \ell + 1}{r} \tpottide[\ell,m,k] 
\end{equation}
evaluated at the surface $r=\rs$, where we introduce
\begin{equation}
  \tpottide[\ell,m,k] \equiv - \epstide \frac{G \Mpri}{\Rpri} \cbar[\ell,m,k] \left( \frac{r}{\Rpri} \right)^{\ell}
\end{equation}
as the radial part of the partial tidal potential~(\ref{e:pottide-part}).
\end{CJK*}
\end{document}

%% file: symbols.tex
\newcommand{\Msun}{M_{\odot}}
\newcommand{\Rsun}{R_{\odot}}

\DeclareSIUnit\dyne{dyn}
\DeclareSIUnit\year{yr}

\newcommand{\Mpri}{M}
\newcommand{\Rpri}{R}
\newcommand{\Lpri}{L}

\newcommand{\mratio}{q}

\newcommand{\vpos}{\mathbf{r}}
\newcommand{\vsec}{\mathbf{r}_{2}}
\newcommand{\rsec}{r_{2}}

\newcommand{\vex}{\mathbf{e}_{x}}
\newcommand{\vey}{\mathbf{e}_{y}}

\newcommand{\ver}{\mathbf{e}_{r}}
\newcommand{\vet}{\mathbf{e}_{\vartheta}}
\newcommand{\vep}{\mathbf{e}_{\varphi}}
\newcommand{\vei}{\mathbf{e}_{i}}

\newcommand{\tanom}{\upsilon}
\newcommand{\eanom}{\mathcal{E}}
\newcommand{\manom}{\mathcal{M}}
\newcommand{\tperi}{t_{0}}

\newcommand{\Oorb}{\Omega_{\rm orb}}
\newcommand{\Porb}{P_{\rm orb}}

\newcommand{\Operi}{\Omega_{\rm peri}}

\newcommand{\epstide}{\varepsilon_{\rm T}}

\newcommand{\sigmk}{\sigma_{m,k}}

\newcommand{\Teff}{T_{\rm eff}}

\newcommand{\Lrad}{L_{\rm R}}
\newcommand{\epsnuc}{\epsilon}

\newcommand{\cP}{c_{P}}
\newcommand{\nablaad}{\nabla_{\rm ad}}
\newcommand{\upsT}{\upsilon_{T}}

\newcommand{\kapad}{\kappa_{\rm ad}}
\newcommand{\kapS}{\kappa_{S}}
\newcommand{\epsad}{\epsilon_{\rm ad}}
\newcommand{\epsS}{\epsilon_{S}}

\newcommand{\Orot}{\Omega_{\rm rot}}
\newcommand{\Opsy}{\Omega_{\rm ps}}

\newcommand{\vFrad}{\mathbf{F}_{\rm rad}}
\newcommand{\Fradr}{F_{{\rm rad},r}}
\newcommand{\vFcon}{\mathbf{F}_{\rm con}}

\newcommand{\tthm}{t_{\rm thm}}

\newcommand{\rs}{r_{\rm s}}

\newcommand{\Slamb}{S_{\ell}}

\newcommand{\potpri}{\Phi}
\newcommand{\potsec}{\Phi_{2}}
\newcommand{\pottide}[1][0]{\Phi_{{\rm T}\if#10\else;#1\fi}}
\newcommand{\pottot}[1][0]{\Psi\if#10\else_{#1}\fi}

\newcommand{\vxi}{\boldsymbol{\xi}}
\newcommand{\xir}{\xi_{\rm r}}

\newcommand{\vv}{\mathbf{v}}

\newcommand{\angtime}[1][0]{H_{\if#10\else#1\fi}}

\newcommand{\txir}[1][0]{\tilde{\xi}_{{\rm r}\if#10\else;#1\fi}}
\newcommand{\txih}[1][0]{\tilde{\xi}_{{\rm h}\if#10\else;#1\fi}}

\newcommand{\tf}[1][0]{\tilde{f}_{\if#10\else#1\fi}}
\newcommand{\tP}[1][0]{\tilde{P}_{\if#10\else#1\fi}}

\newcommand{\tpotpri}[1][0]{\tilde{\potpri}_{\if#10\else#1\fi}}
\newcommand{\tpottot}[1][0]{\tilde{\Psi}_{\if#10\else#1\fi}}
\newcommand{\tpottide}[1][0]{\tilde{\Phi}_{{\rm T}\if#10\else;#1\fi}}

\newcommand{\trho}[1][0]{\tilde{\rho}_{\if#10\else#1\fi}}
\newcommand{\tS}[1][0]{\tilde{S}_{\if#10\else#1\fi}}
\newcommand{\tT}[1][0]{\tilde{T}_{\if#10\else#1\fi}}

\newcommand{\tFradr}[1][0]{\tilde{F}_{{\rm rad},r\if#10\else;#1\fi}}

\newcommand{\teps}[1][0]{\tilde{\epsilon}_{\if#10\else#1\fi}}
\newcommand{\tkap}[1][0]{\tilde{\kappa}_{\if#10\else#1\fi}}
\newcommand{\tLrad}[1][0]{\tilde{L}_{{\rm R}\if#10\else;#1\fi}}

\newcommand{\cbar}[1][0]{\bar{c}_{\if#10\else#1\fi}}
\newcommand{\Fbar}[1][0]{\bar{F}_{\if#10\else#1\fi}}
\newcommand{\Gbar}[2][0]{\bar{G}^{(#2)}_{\if#10\else#1\fi}}
\newcommand{\gbar}[1][0]{\bar{\gamma}_{\if#10\else#1\fi}}

\newcommand{\kap}[1][0]{\kappa_{\if#10\else#1\fi}}

\newcommand{\torque}{\mathcal{T}}

\newcommand{\torsec}{\torque_{\rm sec}}

\newcommand{\tcirc}{t_{e}}

\newcommand{\detune}{\epsilon}
\newcommand{\ndetune}{\bar{\detune}}

\newcommand{\overlap}{Q_{n,\ell}}

\newcommand{\gmode}[1]{{\rm g}_{#1}}

\newcommand{\signl}{\sigma_{n,\ell}}
\newcommand{\gamnl}{\gamma_{n,\ell}}

\newcommand{\mS}{\mathsf{\mathbf{S}}}
\newcommand{\vu}{\mathbf{u}}
\newcommand{\vb}{\mathbf{b}}

\newcommand{\real}{\operatorname{Re}}
\newcommand{\imag}{\operatorname{Im}}

\newcommand{\ee}[1]{\operatorname{e}^{#1}}
\newcommand{\ii}{\mathrm{i}}

\newcommand{\diff}[1]{\operatorname{d}\!{#1}}
\newcommand{\deriv}[3][0]{\frac{\operatorname{d}\if#10\else^{#1}\!\fi\!{#2}}{\operatorname{d}\!{#3}\if#10\else^{#1}\fi}}
\newcommand{\sderiv}[3][0]{\operatorname{d}\if#10\else^{#1}\fi\!{#2}/\operatorname{d}\!{#3}}
\newcommand{\pderiv}[3][0]{\frac{\partial\if#10\else^{#1}\!\fi{#2}}{\partial{#3}}}
\newcommand{\spderiv}[3][0]{\partial\if#10\else^{#1}\fi{#2}/\partial{#3}}

\newcommand{\gyre}{GYRE}
\newcommand{\mesa}{MESA}
\newcommand{\cafein}{CAFein}

\newcommand{\xgyre}{\texttt{gyre}}
\newcommand{\xgyretides}{\texttt{gyre\_tides}}

\newcommand{\BV}{Brunt-V\"ais\"al\"a}
